**The first confirmed superoutburst of the dwarf nova GALEX J215818.5+241924**

Jeremy Shears, Robert Koff, Gianluca Masi, Enrique de Miguel, Ian Miller, George Roberts, Richard Sabo, William Stein and Joseph Ulowetz

**Abstract**

In 2011 October an optical transient was reported in Pegasus as a possible nova. The object had an ultraviolet counterpart, GALEX J215818.5+241924. In this paper we present follow-up photometry of the object which revealed the presence of superhumps, with peak-to-peak amplitude of up to 0.22 magnitudes, diagnostic of it being a member of the SU UMa family of dwarf novae. The outburst amplitude was 4.6 magnitudes and it lasted at least 10 days, with a maximum brightness of magnitude 14.3. We determined the mean superhump period from our first 5 nights of observations as $P_{sh}$ = 0.06728(21) d. However analysis of the O-C residuals showed a dramatic evolution in $P_{sh}$ during the outburst. During the first part of the plateau phase the period increased with $dP_{sh}/dt$ = +2.67(15) x $10^{-4}$. There was then an abrupt change following which the period decreased with $dP_{sh}/dt$ = -2.08(9)$\times 10^{-4}$. We found a signal in the power spectrum of the photometry which we tentatively interpret as the orbital signal with  $P_{orb}$ = 0.06606(35) d. Thus the superhump period excess was $\varepsilon$ = 0.020(8), such value being consistent with other SU UMa systems of similar orbital period.

**Introduction**

The Central Bureau of Astronomical Telegrams (CBAT) Transient Object Confirmation Page listed a possible nova PNV J21581852+2419246 in Pegasus observed by Guoyou Sun and Xing Gao on 2011 Oct 27.6436 at an unfiltered magnitude of 14.5 (1).

Denisenko (2) pointed out that the object has an ultraviolet counterpart GALEX J215818.5+241924 with a far UV magnitude FUV=18.28+/-0.06 and a blue counterpart GSC2.3 N2M9027659 with J=19.13, V=18.86, N=19.10.  He further noted that there are no previous outbursts on 36 NEAT (Near-Earth Asteroid Tracking) images.

The authors commenced a campaign of time-resolved photometry on the object a little over 12 hours after the discovery and the results are reported in this paper

**Photometry and analysis**

We conducted a total of 89 hours of unfiltered time-resolved photometry during the outburst, using the instrumentation shown in Table 1 and according to the observation log in Table 2. Images were dark-subtracted and flat-fielded prior to being measured using differential aperture photometry relative to V-magnitude comparison stars on BAA VSS provisional chart P111028. Given that each observer used slightly different instrumentation, including CCD cameras with different spectral responses, small systematic differences are likely to exist between observers. Where overlapping datasets were obtained, we aligned measurements by different observers by experiment. Adjustments of up to 0.09 magnitudes were made. However, since the aim of the time resolved photometry was to investigate periodic variations in the light curve, we consider this not to be a significant disadvantage. Heliocentric corrections were applied to all data.

## Outburst light curve

The overall light curve of the outburst is shown in the top panel of Figure 1. The object was apparently already in full outburst when first detected (maroon square in Figure 1). This plateau phase was observed for about 8 days, during which the object gradually faded at a mean rate of 0.13 mag/d before it went into rapid decline. At its brightest it was about magnitude 14.3, and taking quiescence as 18.9 (GSC 2.3.2) represents an outburst amplitude of 4.6 magnitudes. The last observation was about 10 days after detection, but at magnitude 17.8 it was still above quiescence. The amplitude and the duration of the outburst are consistent with GALEX J215818.5+241924 being a dwarf nova rather than a nova.

## Measurement of the superhump period

Expanded views of some of the longer time series photometry runs are plotted in Figure 2, having de-trended the data and subtracting the mean magnitude. This clearly shows the presence of regular modulations which we interpret as superhumps. The presence of superhumps is diagnostic that GALEX J215818.5+241924 is a member of the SU UMa family of dwarf novae, making this the first confirmed superoutburst of the star.

To study the superhump behaviour, we first extracted the times of each sufficiently well-defined superhump maximum using the Kwee and van van Woerden method (3) in *Peranso v2.5* (4). Times of 46 superhump maxima were found and are listed in Table 3. An analysis of the times of maximum for cycles 0 to 71 (JD 2455862 to 2455867) assuming an unweighted linear fit allowed us to obtain the following superhump maximum ephemeris:

$$HJD_{max} = 2455862.69705(34) + 0.06728(21) \times E \qquad \text{Equation 1}$$

This gives the mean superhump period for the first five nights of the superoutburst of $P_{sh} = 0.06728(21)$ d. The O–C residuals for the superhump maxima for the complete outburst relative to the ephemeris are shown in the middle panel in Figure 1.

## Superhump evolution

The O-C diagram shows that the superhump period changed significantly during the outburst. Following the scheme of Kato *et al.* (5), who studied the superhump period evolution in a large number of SU UMa systems and found three distinct stages (A,B and C), we interpret the O-C diagram as covering stages B and C. The interval between JD 2455862 and 2455866 (red data points in Figure 1, middle panel) corresponds to Stage B, during which we found an increase in the superhump period with $dP_{sh}/dt = +2.67(15) \times 10^{-4}$ by fitting a quadratic function to the data (red dotted line in Figure 1). There was then a change in period at around JD 2455866-7, following which the period decreased with $dP_{sh}/dt = -2.08(9) \times 10^{-4}$ during stage C as shown by the quadratic fit to the data between JD 2455867 and JD 2455871 (green dotted line in Figure 1). The period transition from stage B to C was sudden, as is common in SU UMa systems including SW UMa, UV Per (5) (6), ASAS J224349+0809.5 (7) and SDSS J073208.11+413008.7 (8). Close inspection of the outburst light curve suggests that the transition in superhump regime appeared to correspond with a temporary slowing in the fading trend during the plateau phase. It is interesting to note that the phase transition in both ASAS J224349+0809.5 and SDSS J073208.11+413008.7 coincided with a similar inflexion in the light curve, which suggests there may actually be a physical change in the accretion disc at this point.

By contrast to what is observed in many SU UMa systems, there was only a modest variation in superhump amplitude during the outburst. The average peak-to-peak amplitude was 0.17 magnitudes. Some superhumps near the beginning of the observed part of the plateau were larger, at around 0.22 magnitudes. There was a slight trend towards a reduction in amplitude in the later stages of the plateau phase.

**Orbital period**

We took the combined time series data between JD 2455862 to 2455867, subtracted the mean magnitude and carried out a Lomb-Scargle analysis. The resulting power spectrum in Figure 3a has as its high peak a signal at 14.84(04) cycles/d (0.06739(18) d) due to the superhumps. Its 1 cycle/d aliases are also present. The error estimates are derived using the Schwarzenberg-Czerny method (9). We then pre-whitened the data with the superhump signal, resulting in the Lomb-Scargle power spectrum in Figure 3b. In this case, the strongest signal is at 15.14(8) cycles/d which we tentatively interpret as the orbital signal, giving $P_{orb} = 0.06606(35)$ d. However, the signal is rather weak and may also be an artefact of the variable superhump period. We therefore suggest that further work should be carried out to confirm our interpretation.

**Estimation of the secondary to primary mass ratio**

Taking our tentative orbital period, $P_{orb} = 0.06606(35)$ d and our mean superhump period of $P_{sh} = 0.06739(18)$ d, both obtained from the Lomb-Scargle analysis, we calculate the superhump period excess $\varepsilon = 0.020(8)$. Such value is consistent with other SU UMa systems of similar orbital period, albeit at the lower end of the range (10).

Patterson *et al.* (11) established an empirical relationship between $\varepsilon$ and q, the secondary to primary mass ratio: $\varepsilon = 0.18*q + 0.29*q^2$. This assumes a white dwarf of ~0.75 solar masses which is typical of SU UMa systems. Our value of ε = 0.020 allows us to estimate q = 0.11.

**Conclusion**

In 2011 October an optical transient was reported in Pegasus and identified as a possible nova being given the preliminary designation PNV J21581852+2419246. The object had an ultraviolet counterpart, GALEX J215818.5+241924. We present follow-up photometry of the object which revealed the presence of superhumps, diagnostic of it being a member of the SU UMa family of dwarf novae. The outburst amplitude was 4.6 magnitudes above quiescence and it lasted at least 10 days with a maximum brightness of magnitude 14.3.

We determined the mean superhump period from our first 5 nights of observations as $P_{sh}$ = 0.06728(21) d, however analysis of the O-C residuals showed a dramatic evolution in $P_{sh}$ during the outburst. During the first part of the plateau phase the period increased with $dP_{sh}/dt$ = +2.67(15) x $10^{-4}$. There was then an abrupt change following which the period decreased with $dP_{sh}/dt$ = -2.08(9)$\times10^{-4}$. The average peak-to-peak amplitude of the superhumps was 0.17 and the maximum 0.222 magnitudes.

Period analysis of the data revealed a signal in the power spectrum, which we tentatively interpret as the orbital signal, giving $P_{orb}$ = 0.06606(35) d. Thus the superhump period excess was ε = 0.020(8), such value being consistent with other SU UMa systems of similar orbital period, albeit at the lower end of the range. From an empirical relationship between ε and the secondary to primary mass ratio, q, we estimate q = 0.11.

Given that this was the first confirmed superoutburst of GALEX J215818.5+241924, we suggest that the object should be monitored for future outbursts to determine its outburst period and the length of its supercycle. Moreover additional photometry during outburst or quiescence (which would necessitate the use of large telescopes, given its faintness) may help to confirm the orbital period.

**Acknowledgements**

The authors gratefully acknowledge the use of SIMBAD and Vizier, operated through the Centre de Données Astronomiques (Strasbourg, France), and the NASA/Smithsonian Astrophysics Data System. WE thank the referees for their helpful comments. We thank Patrick Wils for helpful discussions during the preparation of this paper and our referees for their constructive comments.

## Addresses

JS: "Pemberton", School Lane, Bunbury, Tarporley, Cheshire, CW6 9NR, UK [bunburyobservatory@hotmail.com]

RK: 980 Antelope Drive West, Bennett, CO 80102, USA [bob@AntelopeHillsObservatory.org]

GM: Center for Backyard Astrophysics (Italy), via Madonna de Loco 47, 03023 Ceccano FR, Italy  [gianluca@bellatrixobservatory.org]

EdM: Departamento de Fisica Aplicada, Facultad de Ciencias Experimentales, Universidad de Huelva, 21071 Huelva, Spain; Center for Backyard Astrophysics, Observatorio del CIECEM, Parque Dunar, Matalascañas, 21760 Almonte, Huelva, Spain [demiguel@uhu.es]

IM: Furzehill House, Ilston, Swansea, SA2 7LE, UK [furzehillobservatory@hotmail.com]

GR: 2007 Cedarmont Dr., Franklin, TN 37067, USA,  [georgeroberts0804@att.net]

RS: 2336 Trailcrest Dr., Bozeman, MT 59718, USA [richard@theglobal.net]

WS: 6025 Calle Paraiso, Las Cruces, NM 88012, USA [starman@tbelc.org]

JU: 855 Fair Ln, Northbrook, IL 60062, USA [Joe700A@gmail.com]

| Observer | Telescope | CCD |
|----------|-----------|-----|
| Koff | 0.25 m SCT | Apogee AP-47 |
| Masi | 0.36 m reflector | SBIG ST8-XME |
| de Miguel | 0.28 m SCT | QSI-516ws |
| Miller | 0.35 m SCT | Starlight Xpress SXVR-H16 |
| Roberts | 0.4 m SCT | SBIG ST-8 |
| Sabo | 0.43 m reflector | SBIG STL-1001 |
| Shears | 0.28 m SCT | Starlight Xpress SXVF-H9 |
| Stein | 0.35 m SCT | SBIG ST10XME |
| Ulowetz | 0.235 m SCT | QSI-583ws |

**Table 1: Equipment used**

| Start date in 2011 UT | Start time JD | End time JD | Duration h | Observer |
|-----------------------|---------------|-------------|------------|----------|
| Oct 28 | 2455862.674 | 2455862.780 | 2.5 | Koff |
| Oct 28 | 2455863.216 | 2455863.337 | 2.9 | Masi |
| Oct 28 | 2455863.244 | 2455863.450 | 4.9 | Shears |
| Oct 28 | 2455863.345 | 2455863.510 | 4.0 | de Miguel |
| Oct 28 | 2455863.362 | 2455863.474 | 2.7 | Miller |
| Oct 29 | 2455863.574 | 2455863.794 | 5.3 | Koff |
| Oct 29 | 2455863.605 | 2455863.728 | 3.0 | Roberts |
| Oct 29 | 2455864.335 | 2455864.518 | 4.4 | de Miguel |
| Oct 30 | 2455864.503 | 2455864.742 | 5.7 | Ulowetz |
| Oct 30 | 2455864.556 | 2455864.706 | 3.6 | Sabo |
| Oct 30 | 2455864.592 | 2455864.787 | 4.7 | Stein |
| Oct 30 | 2455864.628 | 2455864.702 | 1.8 | Koff |
| Oct 30 | 2455865.279 | 2455865.523 | 3.5 | de Miguel |
| Oct 31 | 2455865.520 | 2455865.772 | 6.0 | Roberts |
| Oct 31 | 2455865.591 | 2455865.768 | 4.2 | Stein |
| Nov 1 | 2455866.581 | 2455866.761 | 4.3 | Stein |
| Nov 1 | 2455866.590 | 2455866.742 | 3.6 | Roberts |
| Nov 1 | 2455866.623 | 2455866.736 | 2.7 | Ulowetz |
| Nov 1 | 2455867.280 | 2455867.480 | 4.8 | Shears |
| Nov 1 | 2455867.337 | 2455867.486 | 3.6 | Miller |
| Nov 3 | 2455868.565 | 2455868.728 | 3.9 | Sabo |
| Nov 3 | 2455869.287 | 2455869.290 | <0.1 | Shears |
| Nov 5 | 2455871.239 | 2455871.253 | 0.3 | Shears |
| Nov 5 | 2455871.306 | 2455871.418 | 2.7 | Miller |
| Nov 6 | 2455871.605 | 2455871.769 | 3.9 | Stein |

**Table 2: Observations log**

| Superhump cycles number | Superhump maximum (HJD) | Uncertainty (d) | 0-C (d) | Superhump amplitude (mag) |
|---|---|---|---|---|
| 0 | 2455862.7015 | 0.0016 | 0.0048 | 0.16 |
| 8 | 2455863.2359 | 0.0008 | 0.0010 | 0.17 |
| 9 | 2455863.3020 | 0.0004 | -0.0002 | 0.22 |
| 9 | 2455863.3031 | 0.0008 | 0.0009 | 0.18 |
| 10 | 2455863.3706 | 0.0004 | 0.0011 | 0.22 |
| 10 | 2455863.3701 | 0.0004 | 0.0006 | 0.22 |
| 10 | 2455863.3698 | 0.0016 | 0.0003 | 0.21 |
| 11 | 2455863.4377 | 0.0028 | 0.0009 | 0.20 |
| 11 | 2455863.4382 | 0.0008 | 0.0014 | 0.18 |
| 11 | 2455863.4394 | 0.0012 | 0.0026 | 0.18 |
| 12 | 2455863.5048 | 0.0008 | 0.0008 | 0.17 |
| 14 | 2455863.6383 | 0.0012 | -0.0003 | 0.19 |
| 15 | 2455863.7066 | 0.0016 | 0.0007 | 0.18 |
| 25 | 2455864.3768 | 0.0016 | -0.0019 | 0.16 |
| 26 | 2455864.4454 | 0.0004 | -0.0006 | 0.18 |
| 27 | 2455864.5123 | 0.0008 | -0.0009 | 0.16 |
| 27 | 2455864.5127 | 0.0020 | -0.0005 | ND |
| 28 | 2455864.5779 | 0.0024 | -0.0026 | 0.14 |
| 29 | 2455864.6471 | 0.0004 | -0.0007 | 0.16 |
| 29 | 2455864.6473 | 0.0012 | -0.0005 | 0.16 |
| 29 | 2455864.6455 | 0.0012 | -0.0023 | 0.18 |
| 29 | 2455864.6453 | 0.0008 | -0.0025 | 0.17 |
| 30 | 2455864.7124 | 0.0008 | -0.0027 | 0.17 |
| 30 | 2455864.7159 | 0.0020 | 0.0008 | 0.15 |
| 31 | 2455864.7795 | 0.0016 | -0.0029 | 0.15 |
| 39 | 2455865.3222 | 0.0004 | 0.0016 | 0.18 |
| 40 | 2455865.3907 | 0.0008 | 0.0028 | 0.17 |
| 41 | 2455865.4565 | 0.0012 | 0.0013 | 0.17 |
| 44 | 2455865.6590 | 0.0012 | 0.0020 | 0.18 |
| 44 | 2455865.6574 | 0.0008 | 0.0004 | 0.16 |
| 45 | 2455865.7256 | 0.0016 | 0.0058 | 0.17 |
| 45 | 2455865.7246 | 0.0008 | 0.0048 | 0.16 |
| 58 | 2455866.5969 | 0.0004 | 0.0025 | 0.19 |
| 59 | 2455866.6646 | 0.0008 | 0.0029 | 0.18 |
| 59 | 2455866.6687 | 0.0020 | 0.0070 | 0.18 |
| 60 | 2455866.7324 | 0.0016 | 0.0034 | 0.18 |
| 60 | 2455866.7361 | 0.0020 | 0.0071 | 0.18 |
| 69 | 2455867.3405 | 0.0016 | 0.0060 | 0.17 |
| 70 | 2455867.4088 | 0.0016 | 0.0070 | 0.18 |
| 70 | 2455867.4074 | 0.0016 | 0.0056 | 0.18 |
| 71 | 2455867.4729 | 0.0008 | 0.0038 | 0.18 |
| 71 | 2455867.4727 | 0.0016 | 0.0036 | 0.18 |
| 88 | 2455868.6141 | 0.0020 | 0.0013 | 0.17 |
| 89 | 2455868.6832 | 0.0016 | 0.0031 | 0.17 |
| 129 | 2455871.3451 | 0.0024 | -0.0262 | 0.14 |
| 134 | 2455871.6827 | 0.0016 | -0.0250 | 0.14 |

**Table 3: Superhump maximum times and amplitudes**

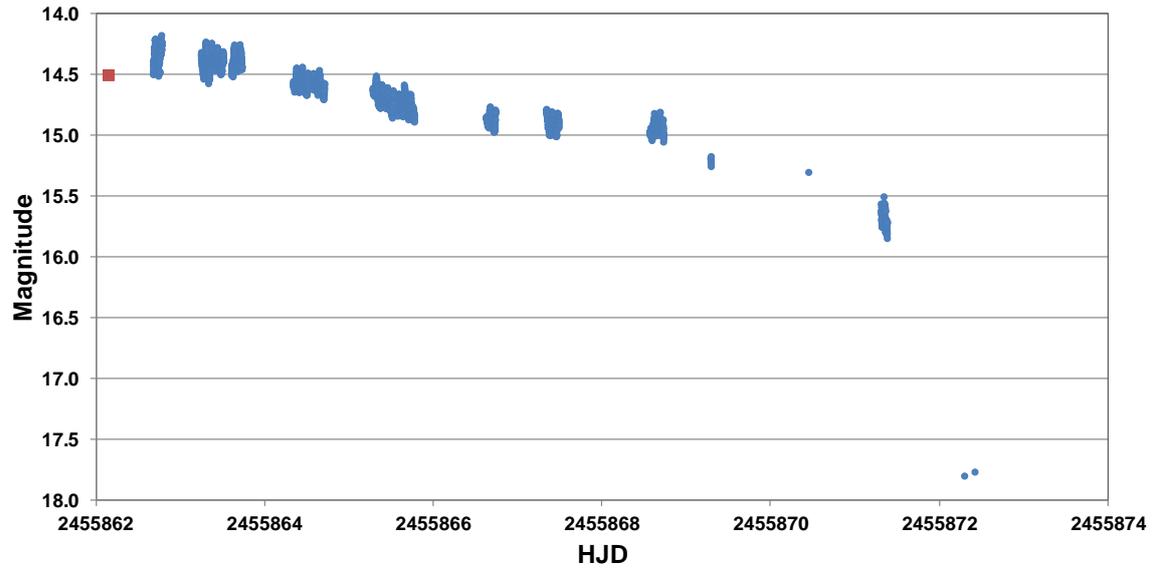

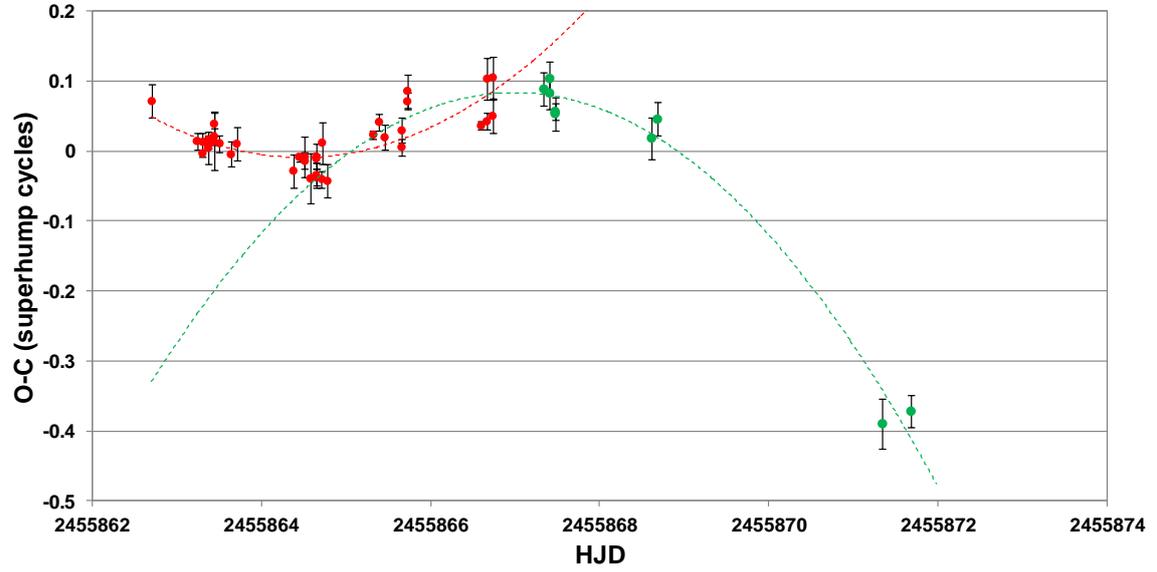

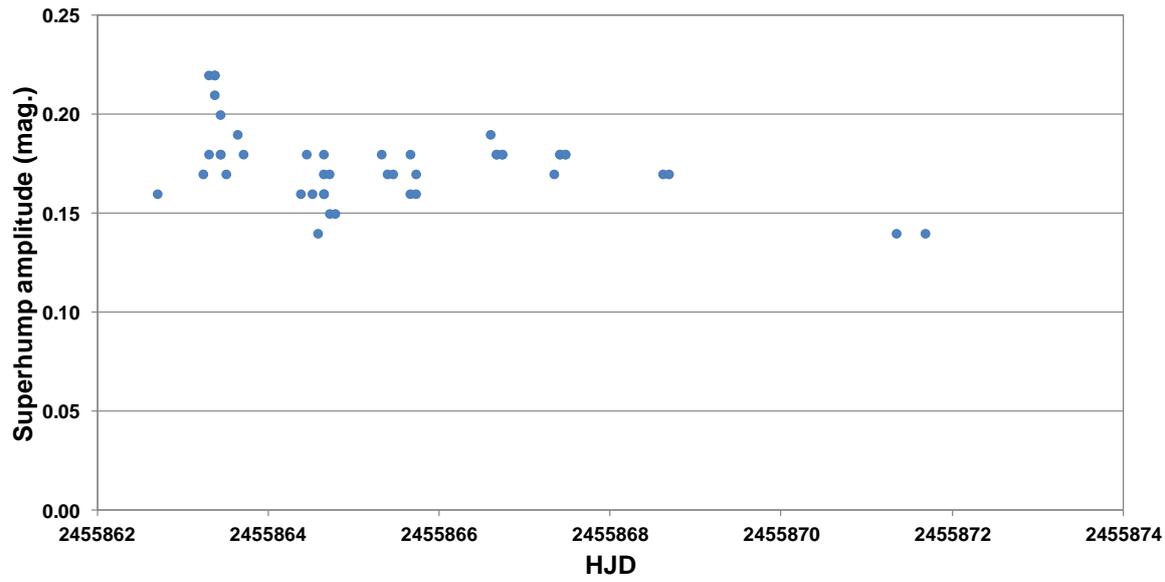

**Figure 1: Light curve of the outburst (top), O-C diagram of superhump maxima relative to the ephemeris in Equation 1 (middle) and superhump amplitude (bottom)**

In the light curve, the CBAT discovery observation is the maroon square – the other data are from the authors. In the O-C diagram, the red dotted line is a quadratic fit to the data between JD 2455862 and 2455866 (red data points) and the green dotted line is a quadratic fit to the data between JD 2445867 to 2445872 (green data points)

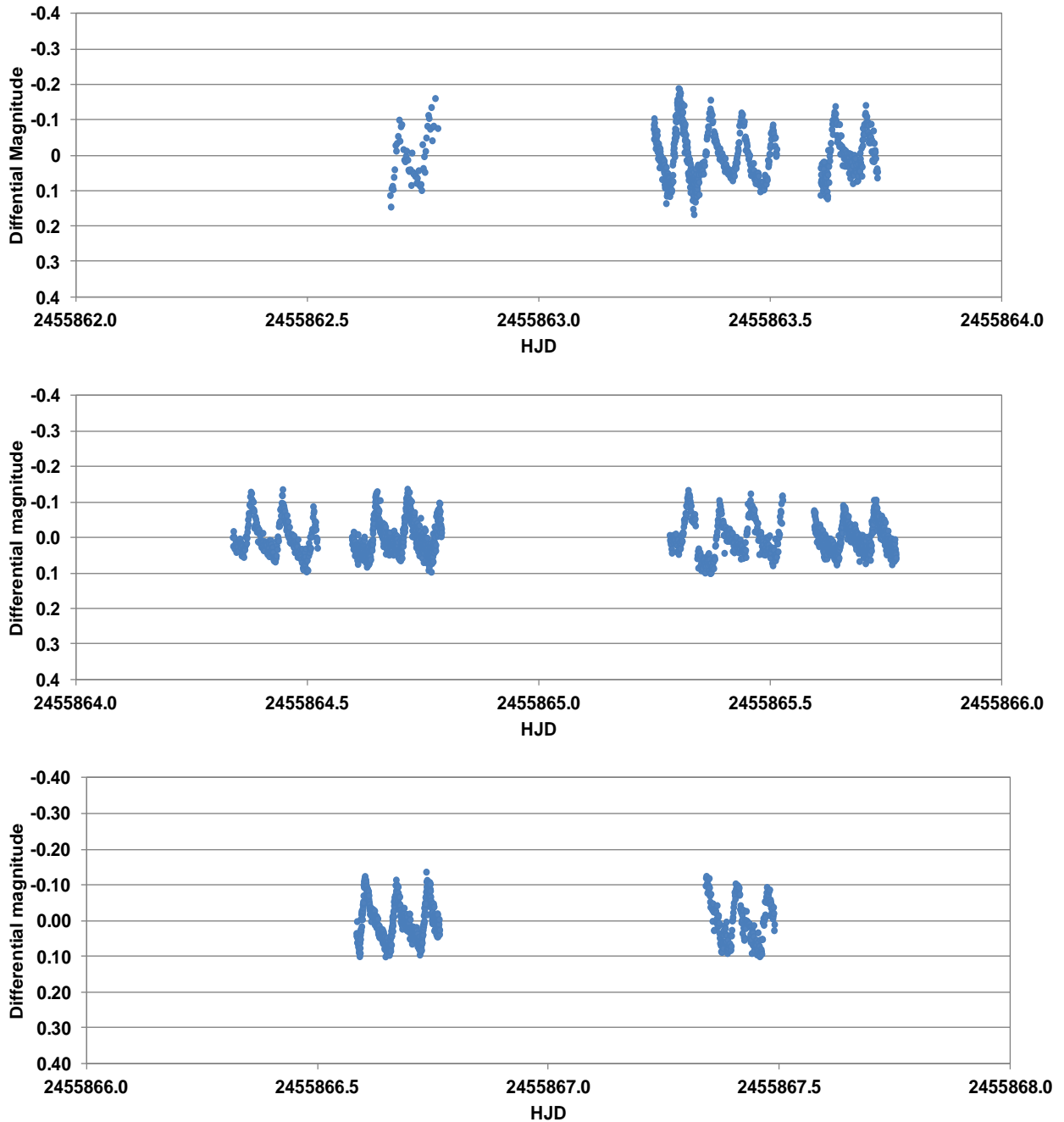

**Figure 2: Time series photometry**

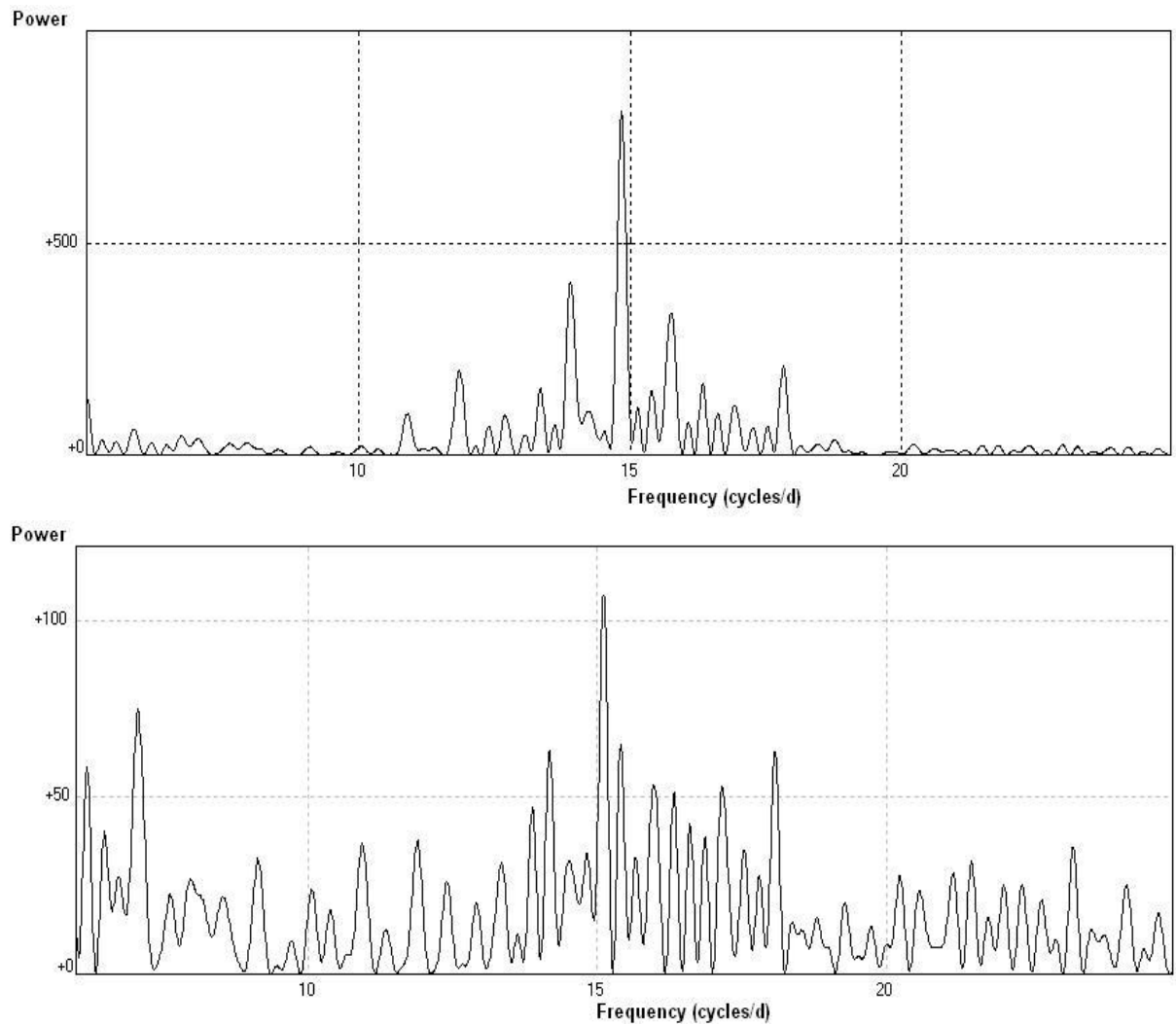

**Figure 3: (a) Lomb-Scargle power spectrum of the time series photometry between JD 2455862 and 2455867, (b) power spectrum after pre-whitening with the superhump signal**